\documentclass{aa}  

\usepackage{graphicx}
\usepackage{txfonts}
\begin{document}

   \title{The protoplanetary system HD 100546 in H$\alpha$ polarized light from SPHERE/ZIMPOL. A bar-like structure across the disk gap? }

 %  \subtitle{I. Overviewing the $\kappa$-mechanism}

   \author{I. Mendigut\'\i{}a\inst{1,2}
          \and
          R.D. Oudmaijer\inst{2}
          \and
          A. Garufi\inst{3}
          \and
          S.L. Lumsden\inst{2}
          \and
          N. Hu\'elamo\inst{1}
          \and
          A. Cheetham\inst{4}
          \and
          W.J. de Wit\inst{5}
          \and
          B. Norris\inst{6}
          \and
          F.A. Olguin\inst{2}
          \and
          P. Tuthill\inst{6}          }          

   \institute{$^{1}$Centro de Astrobiolog\'{\i}a, Departamento de Astrof\'{\i}sica (CSIC-INTA), ESA-ESAC Campus, P.O. Box 78, 28691 Villanueva de la Ca\~nada, Madrid, Spain.\email{imendigutia@cab.inta-csic.es}\\
              $^{2}$School of Physics and Astronomy, University of Leeds, Woodhouse Lane, Leeds LS2 9JT, UK.\\
              $^{3}$Universidad Aut\'onoma de Madrid, Dpto. F\'{\i}sica Te\'orica, M\'odulo 15, Facultad de ciencia, Campus de Cantoblanco, E-28049, Madrid, Spain.\\
              $^{4}$Observatoire de Genève, Université de Genève, 51 chemin des Maillettes, 1290 Versoix, Switzerland\\
              $^{5}$European Southern Observatory, Casilla 19001, Santiago 19, Chile\\
              $^{6}$Sydney Institute for Astronomy (SIfA), Institute for Photonics and Optical Science (IPOS), School of Physics, University of Sydney, NSW 2006, Australia.\\
              }

   \date{Received May 9, 2017; accepted October 26, 2017}

% \abstract{}{}{}{}{} 
% 5 {} token are mandatory
 
  \abstract
   {HD 100546 is one of the few known pre-main-sequence stars that may host a planetary system in its disk.}
   {This work aims to contribute to our understanding of HD 100546 by analyzing new polarimetric images with high spatial resolution.}
   {Using VLT/SPHERE/ZIMPOL with two filters in H$\alpha$ and the adjacent continuum, we have probed the disk gap and the surface layers of the outer disk, covering a region $<$ 500 $mas$ ($<$ 55 $au$ at 109 pc) from the central star, at an angular resolution of $\sim$ 20 $mas$.}
   {Our data show an asymmetry: the SE and NW regions of the outer disk are more polarized than the SW and NE regions. This asymmetry can be explained from a preferential scattering angle close to 90$\degr$ and is consistent with previous polarization images. The outer disk in our observations extends from 13 $\pm$ 2 to 45 $\pm$ 9 $au$, with a position angle and inclination of 137 $\pm$ 5$\degr$ and 44 $\pm$ 8$\degr$, respectively. The comparison with previous estimates suggests that the disk inclination could increase with the stellocentric distance, although the different measurements are still consistent within the error bars. In addition, no direct signature of the innermost candidate companion is detected from the polarimetric data, confirming recent results that were based on intensity imagery. We set an upper limit to its mass accretion rate $<$ 10$^{-8}$ M$_{\odot}$ yr$^{-1}$ for a substellar mass of 15M$_{Jup}$. Finally, we report the first detection ($>$ 3$\sigma$) of a $\sim$ 20 $au$ bar-like structure that crosses the gap through the central region of HD 100546.}
   {In the absence of additional data, it is tentatively suggested that the bar could be dust dragged by infalling gas that radially flows from the outer disk to the inner region. This could represent an exceptional case in which a small-scale radial inflow is observed in a single system. If this scenario is confirmed, it could explain the presence of atomic gas in the inner disk that would otherwise accrete on to the central star on a timescale of a few months/years, as previously indicated from spectro-interferometric data, and could be related with additional (undetected) planets.}

\titlerunning{Protoplanetary system HD 100546 in H$\alpha$ polarized light from SPHERE/ZIMPOL}
   \maketitle
%
%-------------------------------------------------------------------

\section{Introduction}
\label{Sect:Intro}
The exponentially increasing number of known main-sequence stars hosting planetary systems sharply contrasts with the number of forming planets in protoplanetary disks that surround young, pre-main-sequence (PMS) stars, with only a handful of detections to date. While there is debate on the specific process and timescale of planet formation, there is consensus that planetary systems should be completely formed before $\sim$ 10 Myr, a timescale within which circumstellar disks are typically dissipated \citep{Haisch01}. Therefore, a planetary accretion rate of roughly 10$^{-10}$ -10$^{-9}$ M$_{\odot}$ yr$^{-1}$ is required to build $\sim$ 1 -10 M$_{Jup}$ planets within that timescale \citep[see, e.g.,][]{Lovelace11,Sallum15}. Theoretical work indicates that protoplanets accrete material from circumplanetary disks, whose observational signatures are potentially crucial to increase the detection rate of forming planets around young stars \citep[see, e.g.,][]{Zhu15}. For instance, accreting companions with subsolar and planetary masses have been detected from the reduced brightness contrast in H$\alpha$ \citep{Bowler14,Close14,Zhou14,Sallum15}, whose emission is widely used as an accretion indicator.

HD 100546 \citep[d = 109 $\pm$ 4 pc from the first Gaia Data Release;][]{
Lindegren16} is the only young star with a potential planetary system still embedded in the circumstellar disk known to date. A first planet (HD 100546b) was observed in the outer disk at $\sim$ 50 $au$ \citep{Quanz13,Quanz15,Currie14}, although in their recent work, \citet{Rameau17} suggest that the H-band source at the location of HD 100546b is a disk feature. A second candidate planet (HD 100546c) at $\sim$ 10 $au$ was initially proposed by \citet{Bouwman03}. Despite the circumstantial evidence supporting the presence of this second companion, an unambiguous detection has not yet been reported \citep[see, e.g.,][]{Acke06,Brittain09,Brittain13,Brittain14,Boccaletti13,Walsh14,Currie15}. Moreover, \citet{Follette17} indicate that the apparent point source at the location of HD 100546c could result from aggressive data processing. The special properties of the HD 100546 system have been described in a series of dedicated papers using different observational techniques and wavelength regimes that provide complementary information \citep{Sicilia-Aguilar16}. We briefly
summarize them below.
 
\textbf{\textup{The central star.}} Optical spectroscopy was used by \citet{Fairlamb15} to derive a stellar mass, radius, temperature, and age typical of a Herbig Ae/Be star: 1.9 $\pm$ 0.1 M$_{\odot}$, 1.5 $\pm$ 0.1 R$_{\odot}$, 9750 $\pm$ 500 $K$, and 7 $\pm$ 1.5 Myr. Variable emission lines in the optical/UV have been interpreted as the signature of accretion/winds operating in HD 100546 \citep{Vieira99,Deleuil04,guimaraes06}. The mass accretion rate derived from the photospheric excess in the near-UV is $\sim$ 10$^{-7}$ M$_{\odot}$ yr$^{-1}$ \citep{Fairlamb15}, which coincides with the accretion rate derived from the Br$\gamma$ emission \citep{Mendi15b}.

\textbf{The inner region ($<$4 $au$).} An inner dust disk extending from $\sim$ 0.2 $au$ was resolved using near-IR continuum interferometry. The outer edge of the inner dust disk is located at a distance that could range between $\sim$ 1 \citep{Mulders13,Panic14} and 4 $au$ \citep{Benisty10,Tatulli11}. There is evidence of atomic low-density gas traced by the [OI]6300 line on similar spatial scales \citep{Acke06}. In fact, Br$\gamma$ spectro-interferometry revealed an additional atomic gas disk in Keplerian rotation that extends inward of the dust sublimation front \citep{Mendi15b}. Both the dust and gas inner disks are consistent with an inclination ($i$) and position angle ($PA$) of $i$ = 33 $\pm$ 11$\degr$; $PA$ = 140 $\pm$ 16$\degr$, the SW region being closer to us and the NE region farther away \citep{Tatulli11,Mendi15b}. Significantly, \citet{Mendi15b} reported that because of the relatively high mass accretion rate of HD 100546, gas has to be replenished in some way in order to keep a stable inner disk mass ($\sim$ 10$^{-8}$ M$_\odot$). The exact mechanism for this replenishment is unknown, although a planet-induced transfer of gas through the gap from the outer to the inner disk is an exciting possibility \citep[see also][]{Rosenfeld14}. 

\textbf{The gap ($\sim$ 4 - 10 $au$).} The transitional nature of HD 100546 was initially proposed by \citet{Bouwman03} from a spectral energy distribution (SED) analysis. The presence of a gap extending up to 10-15 $au$ has been confirmed from mid-infrared nulling interferometry \citep{Liu03}, spectroscopy in the UV and the near-IR \citep{Grady05,Brittain09,vanderplas09}, mid-infrared interferometry \citep{Panic14}, and high-resolution polarimetric imaging in the optical and near-IR \citep{Quanz11,Garufi16,Follette17}. HD 100546c would be located at some point between the outer part of the gap and the innermost region of the outer disk.  

\textbf{The outer region ($>$ 10 $au$).} Molecular gas traced by CO, OH and CH$^+$ accumulates mainly in the outer disk after the gap, at radial distances from $>$ 10 $au$ up to $\sim$ 400 $au$ \citep{vanderplas09,Panic10,Thi11,Liskowsky12,Hein14}. While $\mu$m dust particles are better mixed with the gas, larger dust particles (mm) do not extend farther than $\sim$ 230 $au$ \citep{Pineda14,Walsh14}. Spiral arms extending beyond $\sim$ 100 $au$ have been detected through high-resolution images in the optical and the near-IR \citep{Grady01,Ardila07,Boccaletti13}. The inclination and position angle of the outer disk have been measured by several groups using various observing techniques, leading to values that are roughly consistent between them \citep[around 45$\degr$ and 145$\degr$ respectively; see, e.g.,][]{Pantin00,Augereau01,Liu03,Ardila07,Panic14,Walsh14}, although several works suggest that the inclination could slightly increase with stellocentric distance \citep{Quillen06,Panic10,Pineda14}. 

Aiming to provide a complementary view of the HD 100546 system, we present new observations taken with the SPHERE/ZIMPOL instrument \citep{Beuzit08,Thalmann08} in polarimetric mode with two filters in H$\alpha$ and the adjacent continuum. This instrumental configuration is ideal to probe the gap and the outer disk of HD 100546 at exquisite angular resolution, also allowing us to explore possible signatures of HD100546c. The wavelength range is also optimal because the amount of reflected light is higher in the optical than in the near-IR \citep[see, e.g.,][]{Draine03,Mulders13}, and the direct, photospheric emission of an A0 star peaks in the optical as well. Section \ref{Sect:Observations} describes the observations and data reduction. The data are analyzed in Sect. \ref{Sect:analisis}, revisiting the properties of the outer disk (Sect. \ref{Sect:geometry}), the candidate companion HD 100546c located at its inner edge (Sect. \ref{Sect:hd100546c}), and the gap that separates the outer and the inner disks (Sect. \ref{Sect:bar}). The summary and conclusions are included in Sect. \ref{Sect:Conclusions}.

\section{Observations and data reduction}
\label{Sect:Observations}
Observations of HD 100546 were taken with SPHERE/ZIMPOL instrument \citep{Beuzit08,Thalmann08} in visitor mode at the European Southern Observatory's Very Large Telescope on 8 June 2015. The average DIMM seeing was 1$\arcsec$, ranging between $\sim$ 0.8$\arcsec$ and 1.2$\arcsec$ during the observing run. The average coherence time was 5 $ms$. Polarimetric observations in pupil-stabilized (P1) ''FastPol'' mode were taken across the full width of the H$\alpha$ emission line shown by HD 100546 \citep{Fairlamb17} and the adjacent continuum with the filters N\_Ha ($\lambda$$_{centr}$ = 656.34 $nm$; $\Delta$$\lambda$ = 0.97 $nm$) and CntHa ($\lambda$$_{centr}$ = 644.90 $nm$; $\Delta$$\lambda$ = 4.1 $nm$), simultaneously. A total of 10 best-seeing polarization cycles out of 12 were selected, each cycle including four frames corresponding to the Stokes components Q$^+$, Q$^-$, U$^+$, and U$^-$. Each single frame had an exposure time of 30 $\times$ 2 $s$ (DIT $\times$ NDIT). The total on-target time was 40 min. 

Data were reduced with the specially developed IDL pipeline at ETH-Z\"urich, which takes care of bias and dark subtraction, flat-field correction, centering, dithering, frame derotation, and combination. The resulting Q (= (Q$^+$ - Q$^-$)/2), U (= (U$^+$ - U$^-$)/2), I (=(I$_Q$ + I$_U$)/2) frames were then used to derive the degree of linear polarization p = (1/I) $\times$ $\sqrt{Q^2 + U^2}$. Standard Q, U frames were used instead of further processing into the more recent Q$_\phi$, U$_\phi$ formalism, given that the former is perfectly valid, especially for disk inclinations similar to or higher than the one of HD 100546 \citep{Canovas15}. Two polarimetric calibration stars were observed during the same night with identical instrumental configuration, and the corresponding data were reduced in the same way. In the optical continuum, HD 111613 is highly polarized \citep[p = 3.1 $\pm$ 0.1 $\%$][]{Bastien88}, whereas HR 5019 (HD 115617) is practically unpolarized \citep[p = 0.010 $\pm$ 0.006$\%$][]{Leroy93}. After inspection of the Q, U frames of the standards, we measured the polarization degree in a resolution element by using a circular aperture with size equal to the SPHERE/ZIMPOL point-spread function (PSF), which for the filters used corresponds to $ \sim$ 20 $mas$ \citep[see also][]{Kervella16}. These measurements are equal in both filters and provide p = 3.7 $\pm$ 0.7 $\%$ for the highly polarized standard and 0.8 $\pm$ 0.4 $\%$ for the unpolarized star. Therefore, the subtraction of a constant value of 0.5 $\%$ - that is, the nominal instrumental polarization of SPHERE/ZIMPOL- from the reduced data of our target suffices to provide a consistent polarization degree within error bars. The calibration images also show that the polarization noise substantially increases from a radial distance of $>$ 140 pixels, which for the SPHERE/ZIMPOL nominal platescale of 3.63 $mas$ pixel$^{-1}$ \citep{Kervella15} corresponds to $>$ 500 mas. Therefore, our analysis of HD 100546 will be focused on smaller angular separations.

\section{Analysis and discussion}
\label{Sect:analisis}
A major advantage of polarimetric differential imaging such as presented here is that direct (unpolarized) light is canceled, allowing us to have access to scattered (polarized) light from dust particles. 

\begin{figure*}
[!hbtp]
\centering
\includegraphics[width=16.8cm,clip=true]{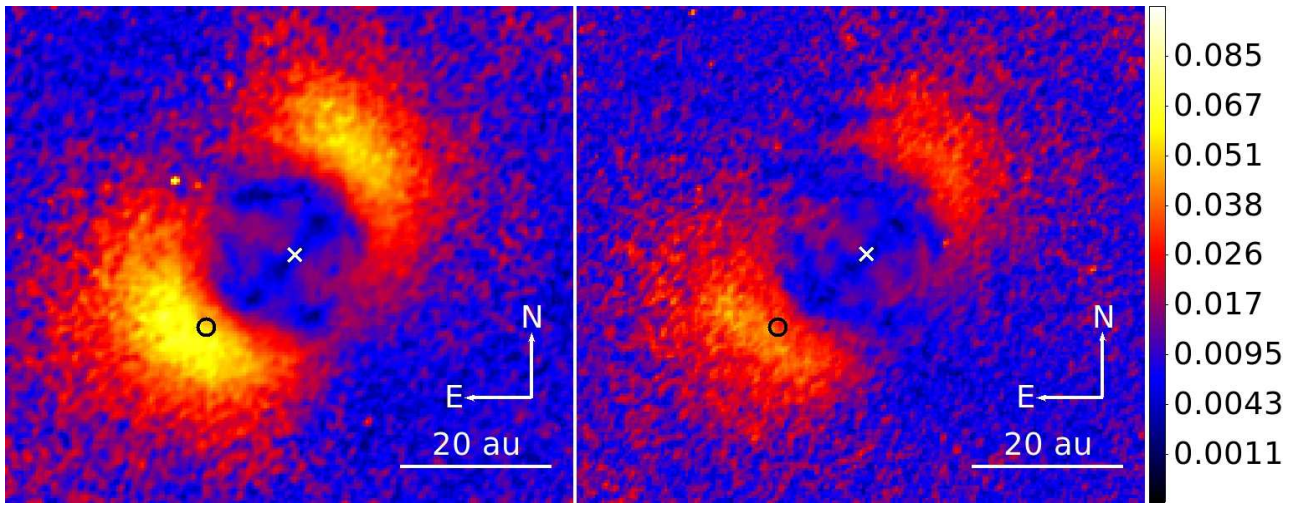}
\caption{SPHERE/ZIMPOL images of HD 100546 in the continuum (left panel) and H$\alpha$ (right panel) filters. The same color code is used for both panels, the numbers in the right-hand legend indicating the linear polarization degree (square root scale). The positions of the central star and candidate c \citep[at a radial distance and $PA$ of $\sim$ 140 $mas$ and 130$\degr$;][]{Currie15} are indicated with a white cross and a black circle, respectively, with a size similar to the spatial resolution ($\sim$ 20 $mas$).}
\label{Figure:images}
\end{figure*}

Figure \ref{Figure:images} shows the degree of polarization of HD 100546 in the continuum and H$\alpha$. The polarization degree is typically lower in H$\alpha$ than in the continuum, which is naturally explained from the fact that the source of the continuum emission comes from a point source (the central star), whereas the source of the H$\alpha$ emission is more extended \citep[accretion/wind regions; see, e.g.,][and references therein]{Mendi15b,Mendi17}. When relatively distant scatterers polarize these two components, the polarization of light coming from a physically extended region (i.e., the Balmer lines) is geometrically diluted compared to the polarization of light arising from a punctual region (i.e., the continuum). Details on this dilution effect have been described by \citet{Cassinelli87} and \citet{Trammell94}, for example. Figure \ref{Figure:images} also shows a clear asymmetry, with the SE and NW regions of the outer disk more polarized in both filters than the SW and NE, as well as a filamentary bar-like structure extending through the gap across the NE-SW direction. The polarization in both filters shows a similar spatial distribution with no clear signature of HD 100546c, whose position is indicated with a black circle. The following sections analyze these observations in more detail.

\subsection{Outer disk}
\label{Sect:geometry}
\begin{figure*}
[!hbtp]
\centering
\includegraphics[width=16.8cm,clip=true]{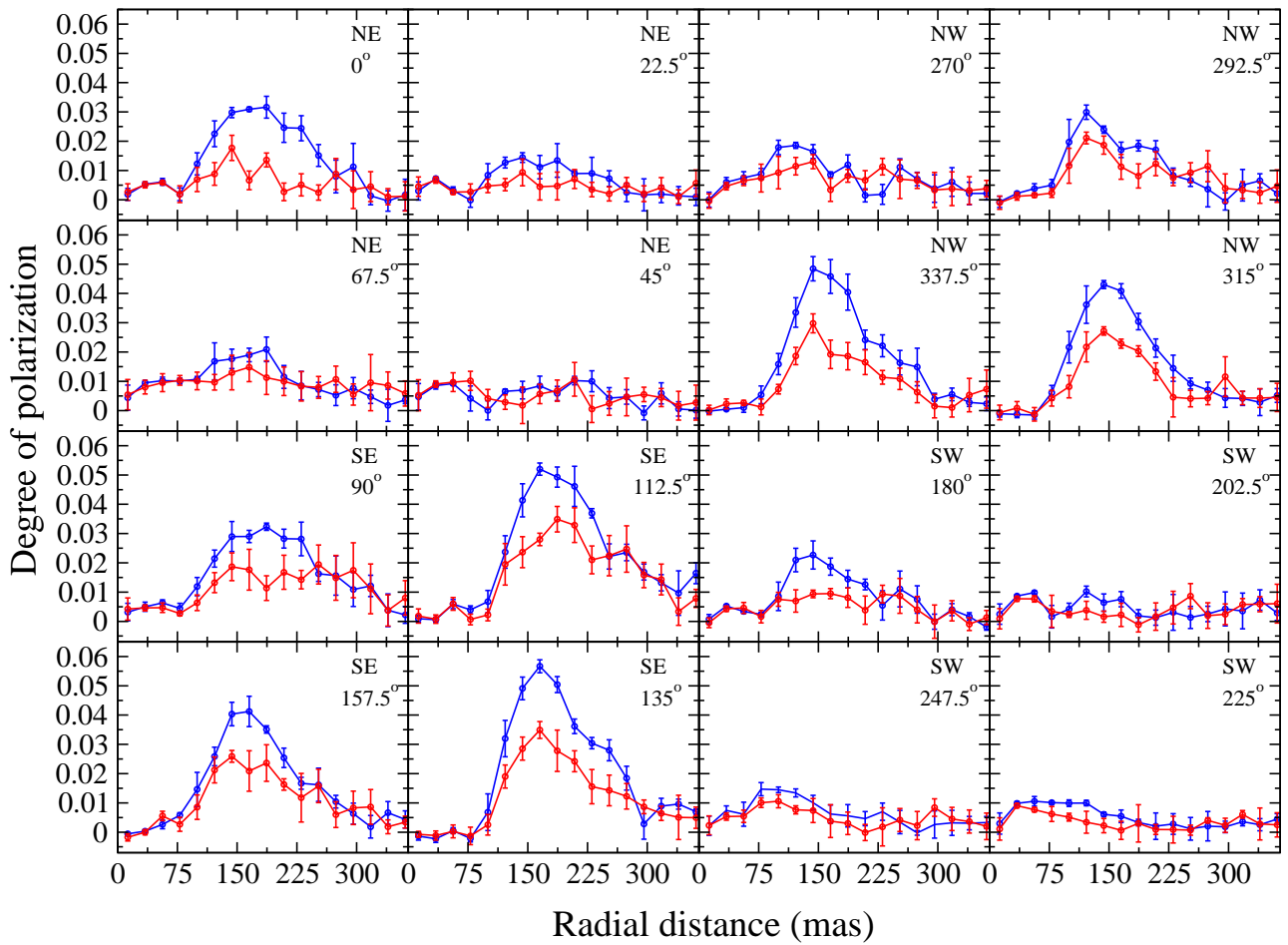}
\caption{Radial profile of the linear degree of polarization for the continuum (blue) and H$\alpha$ (red) filters for the NE (top left panels), NW (top right), SW (bottom right), and SE (bottom left) regions. The $PA$s are indicated in each panel, increasing clockwise in steps of 22.5$\degr$. Each data point is the average of six adjacent pixels along the corresponding $PA$ vector, and the error bars are the corresponding standard deviations.} 
\label{Figure:rad_azim_prof}
\end{figure*}

The disk asymmetry shown in Fig. \ref{Figure:images} can be better quantified from Fig. \ref{Figure:rad_azim_prof}, which shows the stellocentric radial profile of the degree of polarization in the NE, SE, SW, and NW regions. The radial profiles have been derived along four $PA$s per region, as indicated in each panel. The SE region is the most polarized (peak levels of $\sim$ 6$\%$), followed by the NW ($\sim$ 5$\%$). In contrast, the polarization remains within noise levels in most of the NE and SW regions. The polarization shows an abrupt increase at $\sim$ 88 $\pm$ 5 $mas$ in both the SE and NW parts, which corresponds to 13 $\pm$ 2 $au$ after it is deprojected with a disk inclination of 44 $\pm$ 8 $\degr$ (see below) at a distance of 109 $pc$. These observations are consistent with previous determinations of the outer disk wall \citep[see, e.g.,][and references therein]{Garufi16}. Scattered light extends up to 293 $\pm$ 15 $mas$ (45 $\pm$ 9 $au$), and it appears slightly more extended in the SE region than in the NE.

\begin{figure}
%[!hbtp]
\centering
 \includegraphics[width=8.5cm,clip=true]{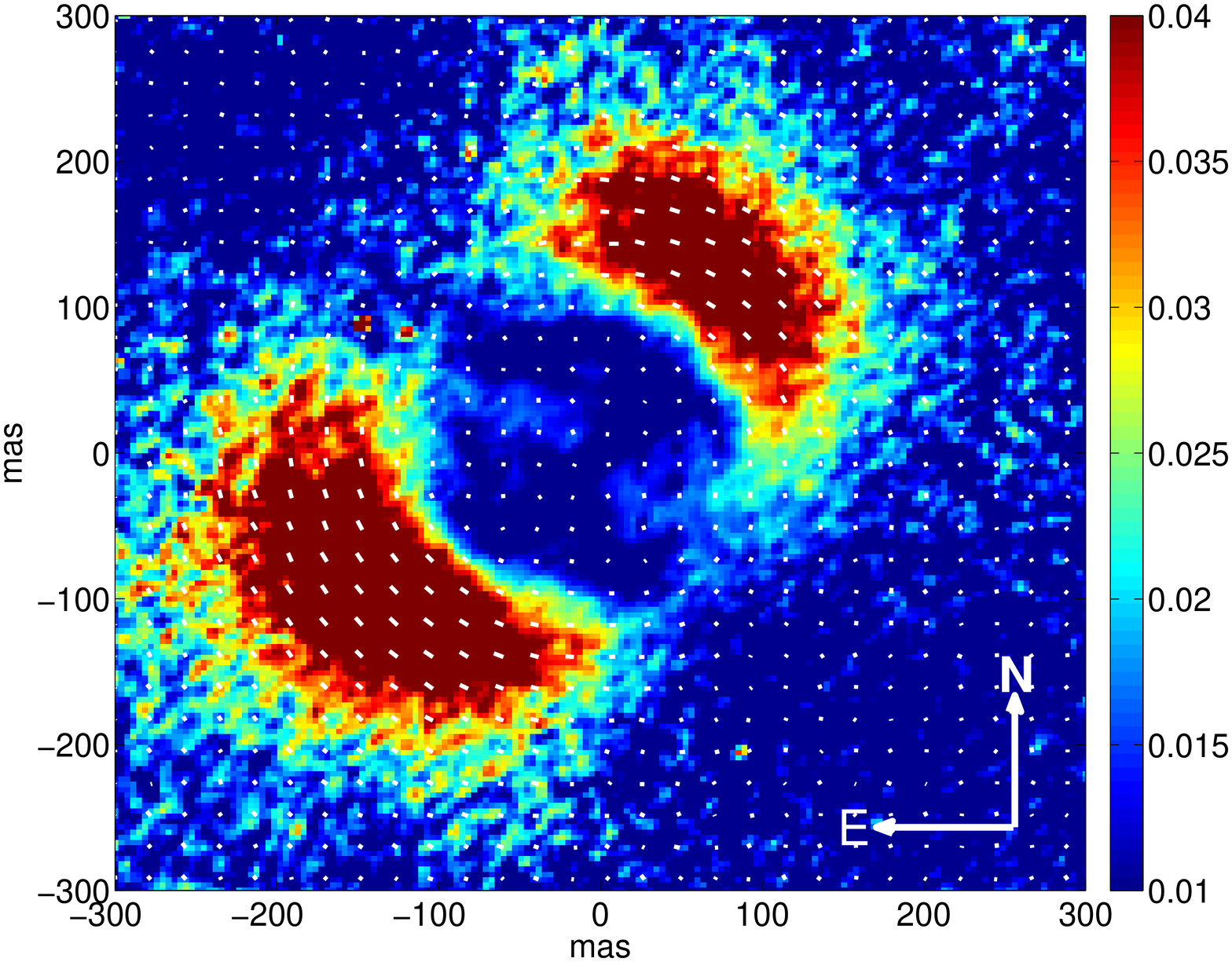}
\caption{White bars indicating the polarization angle (size proportional to the degree of polarization) are overplotted to the continuum image of HD 100546. The numbers in the right-hand legend indicate the linear polarization degree (linear scale).}
\label{Figure:pol_map}
\end{figure}

The polarization asymmetry in the continuum and H$\alpha$ filters was also observed with the SPHERE/ZIMPOL R' filter \citep{Garufi16} and with the NACO H, K, and L' filters \citep{Avenhaus14}. As explained in these works \citep[see also the related discussion and the HD 100546 disk model in][]{Augereau99,Currie15}, an asymmetry like this results from the fact that the polarization efficiency of scatterers is typically maximized for scattering angles around 90$\degr$ with respect to the major axis of the system \citep[e.g.,][]{Perrin09,Murakawa10}. Figure \ref{Figure:pol_map} shows that the angle of linear polarization (0.5 $\times$ tan$^{-1}$(U/Q) + $\alpha$$_0$) in the SE and NW regions is indeed perpendicular to the $PA$ of the major axis of HD100546, aligned in the SE-NW direction (see below). In contrast, forward and backward scattering is minimized, which explains why the SW and NE regions appear to be less polarized.

\begin{figure}
%[!hbtp]
\centering
 \includegraphics[width=8.5cm,clip=true]{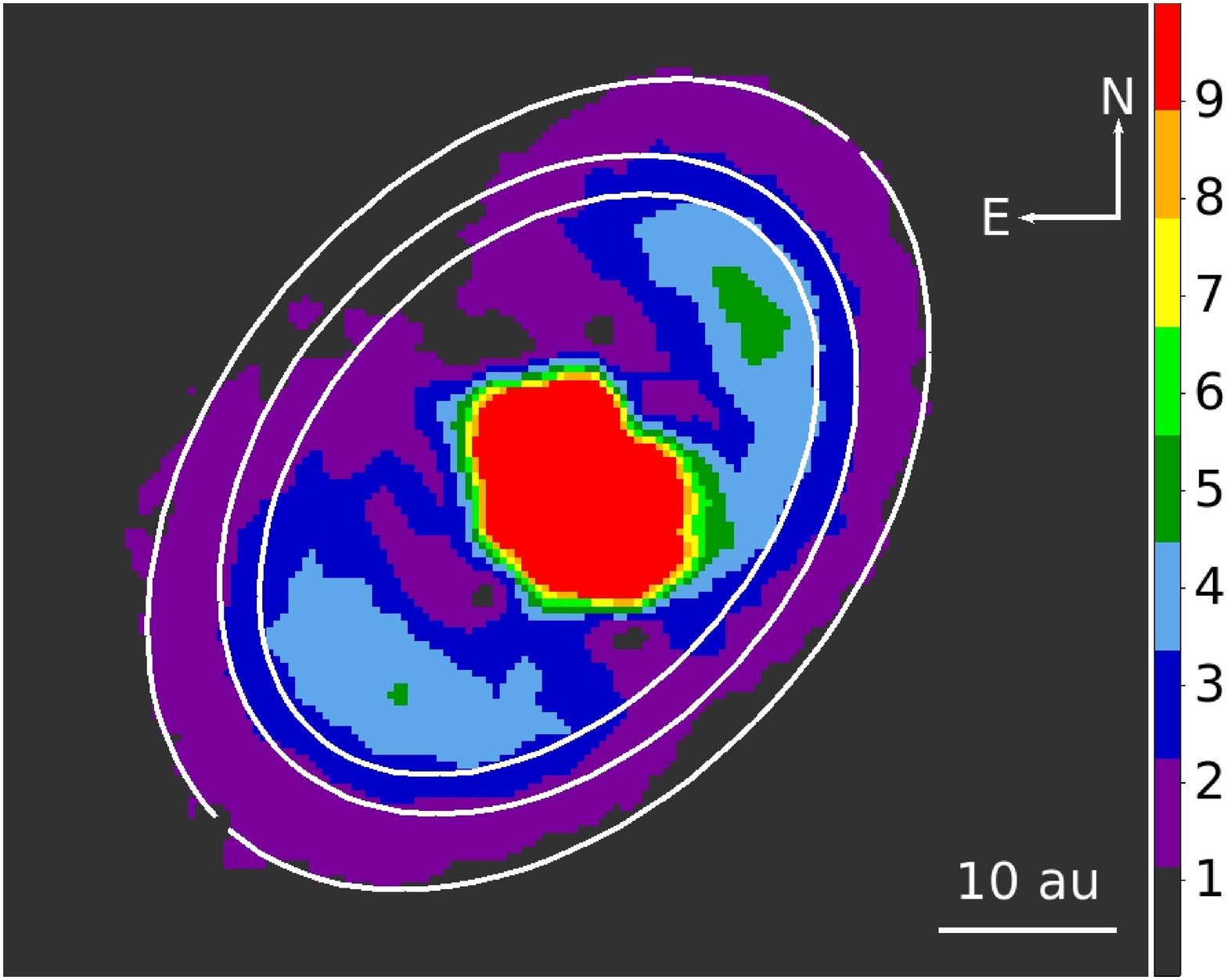}
\caption{Continuum-polarized flux as indicated in the color bar (arbitrary units). Examples of isophotal ellipses are overplotted at a projected radial distance of $\sim$ 26, 29, and 36 $au$.}
\label{Figure:ellipses}
\end{figure}

Several isophotal ellipses were fitted to the continuum polarized flux (= p $\times$ I, Fig. \ref{Figure:ellipses}). Their values for the orientation, major, and minor axes allow us to measure an independent value for the position angle and inclination of the disk. We derive $PA$ = 137 $\pm$ 5$\degr$ and $i$ = cos$^{-1}$(minor axis/major axis) = 44 $\pm$ 8$\degr$ at a radial distance $\sim$ 25-35 $au$. These measurements are consistent with previous derivations of the geometry of the outer disk, which were based on measurements at longer distances. For instance, \citet{Quanz11} derived $PA$ = 138 $\pm$ 4 $\degr$ and $i$ = 47 $\pm$ 3 $\degr$ from isophotal fitting at a radial distance up to $\sim$ 50 $au$. As indicated in that work, all previous direct measurements of $i$ and $PA$ were based on observations at even longer distances from the star. Our results are also consistent with more recent estimates \citep[e.g.,][]{Avenhaus14,Panic14}. All these refer to the first
several $au$ of the outer disk, suggesting that this might be slightly more inclined than the inner disk \citep[$i$ = 33 $\pm$ 11 $\degr$ at a radial distance $\sim$ 0.25 $au;$][]{Tatulli11}, but slightly less inclined than the very outer regions \citep[$i$ $\sim$ 50 $\degr$ at a radial distance $>$ 200 au][]{Quillen06,Panic10,Pineda14}. However, the overlap between the different error bars prevents us from an unambiguous conclusion on the possible change of the inclination with stellocentric distance \citep[see also the recently accepted paper][]{Walsh17}. The analysis of possible relations between this type of a ''warped'' disk with planet formation and other physical processes is beyond the scope of this work \citep[see, e.g., the recent review on the topic in][]{Casassus16}.

\subsection{Constraints on HD 100546c}
\label{Sect:hd100546c}

In this section we derive an upper limit for the accretion rate of HD 100546c, based on the expected influence that the presence of an accreting planet would cause in the ratio between the polarization degree in the continuum and H$\alpha$ filters. In order to do this, we need to express the degree of linear polarization measured in each filter as a fraction $f$ of the corresponding luminosities impinging on the dust grains in the circumstellar/circumplanetary environment; L$_c^*$, L$_{H\alpha}^*$, L$_c^p$, and L$_{H\alpha}^p$ designating the continuum and H$\alpha$ luminosities of the central star and the candidate planet, respectively. The value of $f$ depends not only on the specific filter, but also on the dust grain properties and orientations. Focusing on the regions with higher levels of polarization (NW and SE, which are indicated with superscripts):
\begin{equation}
\label{1}
 p_c^{NW} = f_c^{NW} \times L_c^*,
\end{equation}
\begin{equation}
\label{2}
p_{H\alpha}^{NW} = f_{H\alpha}^{NW} \times L_{H\alpha}^*,
\end{equation}
\begin{equation}
\label{3}
 p_c^{SE} = f_c^{SE} \times (L_c^* + L_c^p) \sim f_c^{SE} \times L_c^*,
\end{equation}
\begin{equation}
\label{4}
p_{H\alpha}^{SE} = f_{H\alpha}^{SE} \times (L_{H\alpha}^* + L_{H\alpha}^p).
\end{equation}
The two former expressions refer to the NW, assuming that the only source of emission in this region is the central star (Eq. \ref{1}) or the accretion/wind regions very close to the stellar photosphere (Eq. \ref{2}). The two latter expressions refer to the SE region, where the companion is supposed to be located. In this case, the contribution of a planetary companion in the optical continuum can be neglected with respect to the star (Eq. \ref{3}), but a possible H$\alpha$ emission resulting from accretion onto the companion could be significant \citep[Eq. \ref{4}; see, e.g.,][]{Bowler14,Zhou14,Sallum15,Zhu15}. For simplicity, it is considered that the fraction of light that becomes polarized in the SE region is roughly the same for both the star and the candidate companion. It is also assumed that L$_{H\alpha}^*$ is isotropic. Equation \ref{4} can be expressed as L$_{H\alpha}^p$ = (p$_{H\alpha}^{SE}$/f$_{H\alpha}^{SE}$) - L$_{H\alpha}^*$. In order to derive an estimate of L$_{H\alpha}^p$, it is finally assumed that the fraction of luminosity that is converted into linear polarization in the SE is proportional to that in the NW, f$_c^{SE}$ = k $\times$ f$_c^{NW}$, and that this proportionality is similar for both filters, f$_{H\alpha}^{SE}$ = k $\times$ f$_{H\alpha}^{NW}$. Combining the two previous expressions with Eqs. \ref{1}, \ref{2}, and \ref{3}, the H$\alpha$ emission from the companion is
\begin{equation}
\label{Eq:Lplanet}
L_{H\alpha}^p = L_{H\alpha}^* \times \left(\frac{p_{H\alpha}^{SE}/p_c^{SE}}{p_{H\alpha}^{NW}/p_c^{NW}} -1\right).
\end{equation}
Equation \ref{Eq:Lplanet} expresses the fact that in the event that a companion is accreting in the SE region of HD 100546, the relative contribution in H$\alpha$ scattered light with respect to the continuum should be stronger in this region than in the NW. In contrast, if both the p$_{H\alpha}^{SE}$/p$_c^{SE}$ and p$_{H\alpha}^{NW}$/p$_c^{NW}$ fractions are similar, L$_{H\alpha}^p$ can only be constrained by an upper limit. It is noted that Eq. \ref{Eq:Lplanet} should be multiplied by a correction factor that depends on the width of the H$\alpha$ filter (Sect. \ref{Sect:Observations}) and on the H$\alpha$ equivalent widths of the central star and the candidate companion. This factor is equal to (1 + FWHM/EW$_*$)/(1 + FWHM/EW$_p$), and serves to address the fact that our measurements not only refer to the H$\alpha$ luminosities, but also to the corresponding continuum luminosities below the emission lines. For the typical H$\alpha$ equivalent widths of HD 100546 \citep[$\sim$ 24 $\AA$;][]{Fairlamb17} and substellar companions detected around other PMS stars \citep[typically hundreds of $\AA$;][]{Bowler14,Zhou14}, the correction factor is close to 1. Given that we are interested in estimating the order of magnitude of the companion's H$\alpha$ luminosity, we consider the current expression in Eq. \ref{Eq:Lplanet} for simplicity.

\begin{figure}
%[!hbtp]
\centering
 \includegraphics[width=8.4cm,clip=true]{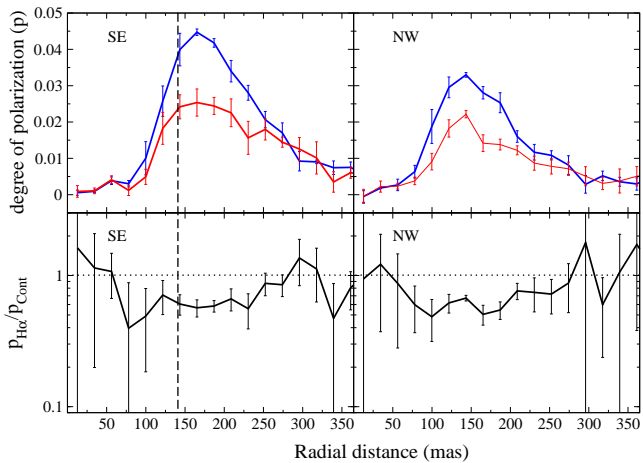}
\caption{(Top) Averaged radial profile of the linear degree of polarization for the continuum (blue) and H$\alpha$ (red) filters as a function of the radial distance in the SE (left, 90$\degr$ $<$ PA $<$ 157.5$\degr$) and NW (right, 270$\degr$ $<$ PA $<$ 315$\degr$) regions. (Bottom) Corresponding ratios (log scale) in the polarization of both filters. The horizontal dotted line indicates equal polarization in both filters and the vertical dashed line the rough radial position of planet candidate c in the SE region.} 
\label{Figure:rad_azim_prof_average}
\end{figure}

The top panels of Fig. \ref{Figure:rad_azim_prof_average} show the averaged degree of linear polarization as a function of the stellocentric distance for the SE and NW regions, derived from the bottom left and top right panels in Fig. \ref{Figure:rad_azim_prof}. The bottom panels of Fig. \ref{Figure:rad_azim_prof_average} show the corresponding fractions of linear polarization from the filters in H$\alpha$ and the adjacent continuum. Although the polarization degree is stronger in the SE than in the NW for both the continuum and H$\alpha$ filters, the relative fractions are similar in both regions. These fractions remain similar when measurements are obtained for smaller regions within the SE and NW parts, with no particular differences at the expected position of HD 100546c. Therefore, no clear signature of an accreting companion is found in the SE region. An upper limit for the companion's H$\alpha$ emission can be derived from Eq. \ref{Eq:Lplanet}. For log (L$_{H_{\alpha}}^*/L_{\odot}$) $\sim$ -1.30 \citep{Fairlamb17} and our polarization measurements and errors, we derive log (L$_{H_{\alpha}}^p/L_{\odot}$) $<$ -3.0. From this value, a limit to the accretion rate can be estimated. The correlation between the H$\alpha$ luminosity and the accretion luminosity is valid for a huge range in mass \citep{Mendi15a,Fairlamb17}. When the correlation for very low mass young stars and brown dwarfs derived by \citet{Herczeg08} is used, the corresponding accretion luminosity is log (L$_{acc}^p/L_{\odot}$) $<$ -1.7. Assuming that the companion has a mass M$_p$ $\sim$ 15M$_{Jup}$ \citep{Boccaletti13,Currie15}, the evolutionary tracks in \citet{Baraffe03} provide a radius R$_p$ $\sim$ 0.13R$_\odot$. Finally, the mass accretion rate can be estimated from the expression $\dot{M}_{\rm acc}$ $\sim$ L$_{acc}^p$R$_p$/GM$_p$, providing an upper limit log (M$_{acc}^p/M_{\odot}$) $<$ -8. 

\citet{Rameau17} and \citet{Follette17} recently carried out a comprehensive study based on high-resolution intensity imagery of HD 100546, deriving upper limits for the accretion rate of the candidate planetary companions. For HD 100546c, even though our estimate for L$_{H_{\alpha}}^p$ is higher than that obtained from simultaneous H$\alpha$ differential imaging in \citet{Follette17}, the upper limit for M$_{acc}^p$ is the same in both works (10$^{-8}$ M$_{\odot}$ yr$^{-1}$). This results mainly from the different assumptions for the planetary mass and radius when L$_{acc}$ is transformed into M$_{acc}$. In particular, \citet{Follette17} used a lower mass (2M$_{Jup}$) and larger radius (0.16R$_\odot$) than in this work. In principle, our intensity frames could also be used to derive an additional estimate for the accretion rate of HD 100546c. However, the reduction techniques and PSF subtraction strongly depend on the field rotation, which was small during our observing run (Sect. \ref{Sect:bar}).

In summary, our polarimetric data confirm that HD 100546c is accreting at $<$ 10$^{-8}$ M$_{\odot}$ yr$^{-1}$, as derived by \citet{Follette17} from intensity data. We conclude that candidate planet c is either in a relatively quiescent stage or its growth from accretion is at a low level or has already ceased \citep{Pinilla15}. However, it is noted that accretion rates $<$ 10$^{-8}$ M$_{\odot}$ yr$^{-1}$ have been measured in other candidate planets and substellar companions \citep{Bowler14,Zhou14,Sallum15}. Alternatively, results from \citet{Follette17} indicate that previous claims on the presence of HD 100546c based on imagery analysis could be related to aggressive data reduction, whereas conservative algorithmic parameters yield a smooth continuous structure consistent with pure disk emission.  

\subsection{Bar within the gap}
\label{Sect:bar}
\begin{figure}
%[!hbtp]
\centering
 \includegraphics[width=8.5cm,clip=true]{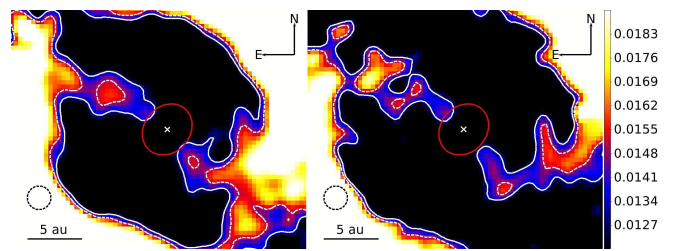}
\caption{Zoom-in of the continuum (left) and H$\alpha$ (right) polarization images showing the bar-like structure extending across NE-SW from the outer to the inner disk (whose outer edge is indicated with the red solid ellipse at $\sim$ 2.5 $au$). The contrast has been varied to highlight the bar, and a Gaussian smoothing of 3 pixels in radius has been applied. +5$\sigma$ and +6$\sigma$ isocontours  above the zero-polarization level are overplotted with white solid and dashed lines. The position of the central star is indicated with a cross and the resolution element with a dashed circle with diameter of 20 $mas$.}
\label{Figure:bar}
\end{figure}

Figure \ref{Figure:bar} shows a zoom-in of the gap. A filamentary bar-like structure is apparent in both the continuum and H$\alpha$ filters, radially extending from the outer to the inner disk NE-SW. The spatial differences in the images derived from the
two filters are very small and smaller than a resolution element. The polarization degree of the bar-like structure is $\sim$ 1/4 that of the most polarized regions of the outer disk, but still consistent with a $>$ 3$\sigma$ detection. The polarization images resulting from the individual Q$^+$, Q$^-$, U$^+$, and U$^-$ cycles of HD 100546 were inspected, and we found a structure
like this in all them. However, this is not present in the standard stars (Sect. \ref{Sect:Observations}).  

\begin{figure}
%[!hbtp]
\centering
 \includegraphics[width=8.5cm,clip=true]{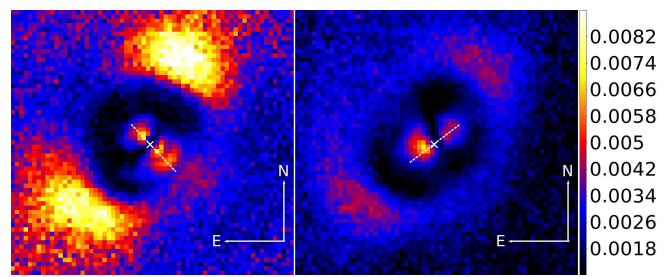}
\caption{HD 100546 from SPHERE-ZIMPOL R'-band observations in P2 ``field-stabilized'' mode from \citet{Garufi16}. The plots show the polarization degree as computed in Sect. \ref{Sect:Observations} (color bar, arbitrary units). The left and right panels correspond to single observations taken at the beginning and at the end of the night, respectively. The field rotation between the two observations was $\sim$ 75$\degr$, which coincides with the rotation of the vector plotted along the diffraction-like artifact that is apparent close to the central star (cross) and inward of the outer disk (which appears more strongly polarized in the left panel).} 
\label{Figure:garufi}
\end{figure}

In addition, we inspected previous images from which the bar could have been observed before. They were taken with SPHERE/ZIMPOL using the R' band in P2 ``field-stabilized'' mode \citep[see][]{Garufi16}. Figure \ref{Figure:garufi} shows that the bar was not apparent in these data. The reason might be that the observing modes were different, where our P1 ``pupil -stabilized'' mode is more efficient than P2 in terms of a higher contrast in regions close to the central star because it minimizes instrumental polarization effects. Our total integration times were also $\sim$ 15 times longer, with individual integration times (DITs) of 30 $s$, against 1.2 $s$ in \cite{Garufi16}. Figure \ref{Figure:garufi} also shows a diffraction-like artifact with the same relative rotation between the field and the spider of the secondary ($\sim$ 75$\degr$). Unfortunately, the field rotation between the first and last cycle in our observations was only $<$ 20$\degr$ , which prevents us from unambiguously ruling out that the bar we have detected is caused by a similar effect as in the R'-band observations. However, it must be noted that not only the bar and the artifact look qualitatively different, but also that the polarization degree of the bar in our observations is much lower than that of the disk and it decreases toward the central star, whereas in the observations from \citet{Garufi16}, the artifact has a polarization degree comparable to the disk and is more polarized closer to the star. Moreover, when the individual cycles in our observations are compared, the bar is visible more clearly as the seeing conditions -and the efficiency of the adaptative optics system- improve. In conclusion, additional non-coronographic SPHERE/ZIMPOL observations in P1 mode with narrow filters that
allow the field to rotate by several dozens of degrees would be very useful to rule out that instrumental effects cause the appearance of the bar-like structure. However, and because of the several lines of evidence provided, we assume hereafter that the bar is a physical feature.

A bipolar outflow with gas sweeping dust from the inner regions along a direction perpendicular to the disk would look similar if it were projected on the sky. However, this type of outflow
is not typical of transitional disks, but is more common in younger
and less evolved stars. In particular, bipolar outflows or Herbig-Haro objects linked to HD 100546 have not been reported before \citep[see, e.g.,][and references therein]{Dent13}. In addition, bipolar outflows and jets tend to be orders of magnitude more extended and do not show evidence of associated dust, which is normally linked to disk winds \citep[see, e.g., the case of the Herbig Ae/Be star HD 163296 in][]{Ellerbroek14}. Moreover, the [OI]6300 emission, normally associated with winds, jets, and outflows, does not reveal any blueshifted component in the case of HD 100546 \citep{Acke05,Acke06}. For these reasons, we consider the outflow hypothesis to be very unlikely.   

Alternatively, the bar could be dust dragged by gas radially flowing from the outer to the inner disk. This type of non-Keplerian flows has been reported before, but it is spread over larger scales and is probably related to binary or multiple systems. \citet{Mayama10} observed a $\sim$ 500 $au$ bridge of infrared emission between two protoplanetary disks in the system \object{SR 24,} which was attributed to a gas stream. Similarly, a gas and dust radial ``streamer'' of $\sim$ 100 $au$ appears to flow from the outer circumbinary ring onto the central region of the multiple system \object{GG Tau} \citep{Pietu11,Beck12}. On a similar scale, a radial inflow crossing the $\sim$ 120-140 $au$ gap and emitting both in the continuum and in molecular lines was claimed by \citet{Casassus13} for the transitional disk \object{HD 142527}, which also has a close stellar companion \citep{Biller12,Close14,Lacour16}. If the radial inflow of \object{HD 100546} were to be confirmed, this would extend over a spatial scale an order of magnitude smaller than in previous reports\footnote{Although ALMA data might be suggesting that a non-Keplerian inflow in HD 100546 may also exist at larger scales than inferred from our observations; \citet{Walsh17}.} and would not be related with any stellar companions. To our knowledge, the only possible analogy could be the classical T Tauri star \object{AA Tau}, for which a $\sim$ 30 $au$ stream connecting the opposite sides of the innermost ring has recently been suggested by \citet{Loomis17} at much closer distances than its candidate companion \citep{Itoh08}.

\citet{Rosenfeld14} proposed that non-Keplerian radial inflows would explain why many transitional disks with wide gaps show similar stellar accretion rates as non-transitional disks, as is the case for HD 100546. Moreover, based on previous spectro-interferometric data, \citet{Mendi15b} reported that the atomic gas in the inner disk is expected to dissapear on a timescale of only a few months
or years, given the relatively high stellar accretion rate of HD 100546. \citet{Mendi15b} suggested that a radial inflow from the outer disk could be replenishing the inner disk of HD 100546, allowing it to keep a stable accretion rate. The bar that we have detected could be a confirmation of this scenario. In this case, the flowing material does not seem to be directly related to either HD 100546b or HD 100546c, as inferred from their distant projected positions. 

Additional observations and modeling are necessary to understand the nature of the bar-like structure. For the former, velocity-resolved observations could provide very helpful information in order to confirm the inflow hypothesis against the outflow scenario. A flow of material from the outer to the inner disk would in principle show velocities close to free-fall \citep{Rosenfeld14}, which for HD 100546 translates into $\sim$ 450-500 km s$^{-1}$ at a radial distance of $\sim$ 6 $au$ from the star, if it were projected with an inclination angle of $\sim$ 44$\degr$. In contrast, the outflow scenario would show lower radial velocities on the order of 200-300 km s$^{-1}$ \citep{Ellerbroek14}. If the inflow hypothesis were to be confirmed, modeling suggests that additional forming planets that have not been detected so far could be located along the radial streams that connect the outer and the inner disk \citep{Artymowicz96,Bryden99,Kley99,Lubow99,Lubow06}. Alternatively, the apparent symmetry between the NE and SW streams resembles a scaled-down version of the bar in some spiral galaxies, whose formation might be explained by dynamical effects without invoking the presence of companions \citep[see, e.g.,][and references therein]{Rosenfeld14}.  

\section{Summary and conclusions}
\label{Sect:Conclusions}
The analysis of new polarimetric data in H$\alpha$ and the adjacent continuum taken with the SPHERE/ZIMPOL instrument at the VLT provides the following results on the protoplanetary system HD 100546:
\begin{itemize}
 \item HD 100564 has an observed outer disk asymmetry in which the SE and NW regions are more strongly polarized than the SW and NE. This asymmetry can be explained from the fact that dust particles have a preferential scattering angle close to 90$\degr$, as it was shown from the analysis of previous polarization images. The outer disk in our observations extends from 13 $\pm$ 2 $au$ to 45 $\pm$ 9 $au$, with a position angle and inclination of 137 $\pm$ 5$\degr$ and 44 $\pm$ 8$\degr$, respectively. The comparison with previous estimates suggests that the disk inclination could increase with the stellocentric distance, although the different measurements are still consistent within error bars.      
 \item There is no signature of the innermost candidate companion HD 100546c. The polarimetric images provide an upper limit for the accretion luminosity of log (L$_{acc}^p/L_{\odot}$) $<$ -1.7, which for a substellar mass of 15M$_{Jup}$ translates into a mass accretion rate $<$ 10$^{-8}$ M$_{\odot}$ yr$^{-1}$, confirming recent estimates that were based on imagery analysis. Our result suggests that either HD 100546c is in a low level of episodic accretion, or that this has already ceased.
 \item A filamentary, radially extended, bar-like structure crossing the gap in the NE-SW direction is detected for the first time in HD 100546. If its physical nature were confirmed, we suggest that the most likely scenario is an inflow of material channelled from the outer disk to the inner region. This would be one of the few cases, if not the only case, in which such a small-scale radial inflow is observed in a system that is not binary or multiple. This scenario might explain the relatively high stellar accretion rate of HD 100546 and could be related with the presence of additional (undetected) planets. 
\end{itemize}
New observations and modeling are necessary to better understand the HD 100546 system. On the one hand, velocity-resolved observations are crucial to confirm the nature of the bar-like structure and distinguish different scenarios. On the other hand, modeling is also necessary to physically understand its origin in possible relation with planetary formation, and how far the analogy with galactic bars and their dynamics holds.

\begin{acknowledgements}
The authors acknowledge the anonymous referee for the useful comments.\\      
The authors thank H.M. Schmid and the ETH group for providing the SPHERE/ZIMPOL pipeline and helping with the data reduction process.\\
The authors also thank C. Walsh for sharing her results on ALMA data before publication.\\ 
IM acknowledges the Government of Comunidad Aut\'onoma de Madrid, Spain, which has partially funded this work through a ``Talento'' Fellowship (2016-T1/TIC-1890).\\
This work is based on observations collected at the European Southern Observatory under ESO programme 095.C-0352(A).\\
This work has made use of data from the European Space Agency (ESA)
mission {\it Gaia} (\url{https://www.cosmos.esa.int/gaia}), processed by
the {\it Gaia} Data Processing and Analysis Consortium (DPAC,
\url{https://www.cosmos.esa.int/web/gaia/dpac/consortium}). Funding
for the DPAC has been provided by national institutions, in particular
the institutions participating in the {\it Gaia} Multilateral Agreement.\\

\end{acknowledgements}

\end{document}